\documentclass[10pt]{iopart}

\usepackage{iopams}
\usepackage{bm}
\usepackage{graphicx}
\usepackage{comment}
\usepackage{epsfig}                  

\renewcommand{\>}{\right \rangle}

\newcommand{\ket}[1]{\left |#1\>}

\newcommand{\be}{\begin{equation}}
\newcommand{\ee}{\end{equation}}
\newcommand{\bea}{\begin{eqnarray}}
\newcommand{\eea}{\end{eqnarray}}

\newcommand{\aver}[1]{\langle #1 \rangle}

\begin{document}
\title[Effects of phase and interactions on atom-laser outcoupling 
from a double-well BEC]{Effects of relative phase and interactions on atom-laser 
outcoupling from a double-well Bose-Einstein condensate: Markovian and 
non-Markovian dynamics}

\author{G. M. Nikolopoulos$^{1}$, C. Lazarou$^{2}$ and P. Lambropoulos$^{1,3}$}

\address{$^1$ Institute of Electronic Structure and Laser, FORTH, P. O. Box 1527, 
Heraklion 711 10, Crete, Greece}

\address{$^2$ Department of Physics and Astronomy, University of Sussex, 
Brighton BN1 9QH, United Kingdom}

\address{$^3$ Department of Physics, University of Crete, P. O. Box 2208, 
Heraklion 71003, Crete, Greece}


\begin{abstract}
We investigate aspects of the dynamics of a continuous atom-laser 
scheme based on the merging of independently formed atomic condensates. 
Our theoretical analysis covers the Markovian as well as the non-Markovian  
operational regimes, and is based on a semiclassical 
(mean-field) two-mode model. 
The role of the relative phase between the two condensates and 
the effect of interatomic interactions on the evolution of the 
trapped populations 
and the distribution of outcoupled atoms are discussed. 

\end{abstract}



\maketitle

\section{Introduction}
The dynamics of the atom laser, beyond the Born and Markov approximations 
have been investigated mostly in the framework of a rather simple model, 
involving a single condensate mode coherently coupled to a continuum of 
free-space modes 
\cite{moypra299,hopepra00,jackpra99,bre99,jefpra00,NLP03}. 
In recent work \cite{LazNikLam07}, we extended this model to a two-mode 
scenario. Specifically, we considered two initially independent Bose-Einstein 
condensates (BECs)  
consisting of a large number of bosonic atoms cooled into the lowest 
eigenmode of their respective traps. 
The two traps are assumed brought together, while atoms are outcoupled 
coherently from one of the traps only, whose condensate mode plays the 
role of the 
``lasing mode'' which is replenished with atoms from the other trap, 
which serves as a ``pump''. 
The chief motivation for this investigation came from 
the experiment of Chikkatur {\rm et al.} \cite{MITscience02} 
on the merging of BECs, which offers a conceptually simple scenario for a 
continuous atom laser. 
The main idea in that work relies on replenishing the condensate mode 
of the lasing trap by transfering other independently formed BECs using 
optical tweezers.

The theoretical modelling of such a scenario has to cope with a reasonable 
description of the condensates, the non-Markovian character of the coupling 
to the atomic continuum, as well as the interatomic interactions present in 
the system. 
In an attempt to understand how the presence of the second condensate affects 
the atom-laser dynamics, we focused first on an interaction-free 
model, assuming in addition a vanishing relative phase 
between the two condensates  \cite{LazNikLam07}. 
Even within these two simplifications, the system was found to exhibit a number 
of interesting features such as the formation of a bound state in the 
strong-outcoupling regime, as well as a dark line in the spectrum of the 
outcoupled atoms. 
In the present paper, we relax both of the above simplifying assumptions,  
addressing thus a more realistic 
scenario which takes into account inter-particle interactions and a non-vanishing 
relative phase between the two independent BECs. As one might have anticipated, 
a number of new features emerge in the time-development of the system as well as 
the spectrum of the outcoupled atoms. 

In brief, the interplay among several parameters such as the changing chemical 
potentials due to the outcoupling, the strength of the interparticle interactions 
(nonlinearity), as well as the strength of the outcoupling, 
set the stage for the appearance of a Josephson effect which competes with 
self-trapping, and also chirping, i.e. a variation of the energies of the 
outcoupled atoms. 

\section{The system}
\label{SecII}
Our system  consists of two independently 
prepared, elongated BECs (A and B), with $N$ being the total number 
of atoms in the system. 
The two traps are initially far apart and each BEC experiences 
only its local potential, while   
only the lowest level of each trap (condensate mode) is populated.
To allow for the merging of the two BECs, trap B is moved towards 
trap A along the radial direction $x$, while at the same time  
atoms are outcoupled coherently from BEC A, by applying external 
electromagnetic fields (see Ref. \cite{LazNikLam07} for a detailed description 
of the system under consideration). 
Hence, to point out the discrete roles of the two BECs, from now on we also 
refer to BEC A as the ``lasing condensate'' and to BEC B as the ``pumping condensate''.  

Presently, transport of BECs can be realized efficiently using optical 
tweezers which are produced by focused laser beams and offer limited trap 
volume and depth \cite{MITscience02}.  
Hence, during the merging process the two BECs can be brought as close 
as the traps' beam waist, before they start affecting each other. 
To be consistent with the experimental setup for BEC merging 
\cite{MITscience02} 
as well as related theoretical work \cite{YiDuanpra05,MeSaLepra06}, 
we assume two nearly identical axially symmetric harmonic traps with confining 
frequencies $\omega_z$ and $\omega_x=\omega_y=\omega_\perp$. The 
atomic density profiles adiabatically follow the movement of the traps,  
while the global potential experienced by the trapped atoms can be 
modeled by a time-dependent double-well potential \cite{LazNikLam07}.
In general, besides the merging time-scale $t_{\rm m}$, the details of the motion 
of trap B are not of great importance 
\cite{MITscience02,YiDuanpra05,MeSaLepra06}. 
The crucial point is that the BEC merging must be 
adiabatic so that any kind of excitations in the system are suppressed. 

The outcoupling mechanism under consideration is based on the application of external 
electromagnetic fields which induce an 
atomic transition from the internal state $(\ket{t})$ of the trapped atoms 
to an untrapped state $\ket{f}$. The outcoupled atoms are guided through 
an atomic waveguide  \cite{book,MHSguide}, thus 
resulting in an effective one-dimensional atom laser propagating 
along the weak confining axis of the waveguide. 
Such a guided atom laser has been demonstrated 
recently by Guerin {\em et al.} \cite{Gueprl06} and offers many advantages 
over the conventional outcoupling schemes. 
Formally speaking, the strong transverse confinement allows us to assume 
that the transverse dynamics of the free atoms adiabatically follow the 
slowly varying transverse potential of the optical guide \cite{Gueprl06}, 
which is assumed to be nearly the 
same with the transverse potential of trap A.
In the absence of gravitational or other forces (as in the experimental 
setup \cite{Gueprl06}), the longitudinal component of the 
potential is practically zero and the outcoupled atoms propagate freely along 
the $z$ direction.

\subsection{Two-mode model}
During the merging, the condensate wavefunctions start overlapping in space 
as the two traps approach each other along the $x$ direction, 
and a Josephson-type tunneling is established between the 
two BECs. If the position uncertainty in the ground state of the 
traps is much smaller than the separation of the minima of the global 
potential, the overlap (and thus the Josephson coupling $J$) is small 
enough so that only the ground 
states of the traps are relevant. In first-order perturbation theory, 
the corresponding local ground-state wavefunctions 
$\varphi_{\rm A(B)}({\bf r})$ are orthogonal and describe faithfully 
BEC A and B, at any time $0<t\ll t_{\rm m}$ \cite{TwoModeValid}. 
If in addition the effect of interatomic interactions on the ground-state 
properties of the two wells is small i.e., if  
\be
(\omega_x\omega_y\omega_z)^{1/3}\gg N\kappa,
\label{int-as}
\ee
where $\hbar\kappa$ is the onsite interaction energy per particle,  
the system under consideration is well-described in the framework of the 
standard two-mode model \cite{TwoModeValid}. 
Introducing the sets of bosonic operators 
$\{\hat{a}^\dag,\hat{a}\}$ and $\{\hat{b}^\dag,\hat{b}\}$ for the 
description of the condensate modes A (lasing) and B (pumping), 
respectively, the Hamiltonian of the system reads \cite{LazNikLam07}
\bea
\fl\hat{\cal H}=\frac{\hbar\omega_z}{2}(\hat{a}^\dag\hat{a}
+\hat{b}^\dag\hat{b})+\hbar\sum_k\omega_k\hat{c}_k^\dag \hat{c}_k
+\hbar \kappa (\hat{a}^\dag\hat{a}^\dag\hat{a}\hat{a}+
\hat{b}^\dag\hat{b}^\dag\hat{b}\hat{b})
+\hbar J(\hat{a}^\dag\hat{b}+\hat{b}^\dag\hat{a})\nonumber\\
\fl+\hbar\sum_k g(k,t) (\hat{a}\hat{c}_k^\dag
+\hat{a}^\dag\hat{c}_k )+
\hbar e^{-\eta^2}\sum_k g(k,t) (
\hat{b}\hat{c}_k^\dag +\hat{b}^\dag\hat{c}_k),
\label{fullham}
\eea
where $\hat{c}_k^\dag(\hat{c}_k)$ is the creation(annihilation) bosonic 
operator of a free atom with momentum $k$, and frequency 
\be
\omega_k=\frac{\hbar k^2}{2m},
\label{disrel}
\ee
with $m$ the atomic mass. 
In deriving Hamiltonian (\ref{fullham}) we have applied the rotating-wave 
approximation 
for the outcoupling Hamiltonian, while we have neglected 
higher-order cross-interaction terms. Moreover, it is worth noting here that 
although the outcoupling mechanism is applied to the lasing BEC only, 
atoms are also outcoupled from the pumping condensate; albeit with a much 
slower rate [compare the last two terms of Hamiltonian (\ref{fullham})]. 
This is due to the overlap of the two BEC wavefunctions and, as we 
will see later on, it may affect considerably the spectrum of the atom laser. 

For a weakly-interacting bosonic gas satisfying (\ref{int-as}), 
the ground-state wavefunctions 
are well approximated by Gaussians and thus analytic expressions for 
all the parameters entering the Hamiltonian are readily 
obtained \cite{LazNikLam07}
\bea
&&J(t)=\omega_z\left (\frac{1}2+\frac{1}{\lambda} 
-\frac{\eta}{\lambda\sqrt{\pi}}\right )e^{-\eta^2},
\label{Jt_eq}
\\
&&\kappa=\frac{\hbar a_{tt}}{\lambda m\sqrt{2\pi}l_z^3},
\label{kap_eq}\\
&&g(k,t) = \frac{\sqrt{l_z}}{\pi^{1/4}}\sqrt{\Lambda(t)} e^{-k^2 l_z^2/2},
\eea
where $\Lambda(t)$ is the coupling between trapped and untrapped atomic states, 
$\eta$ is the dimensionless separation of the traps, $\lambda=\omega_z/\omega_x$, 
and $l_z\equiv\sqrt{\hbar/m\omega_z}$.
In view of these relations, condition  (\ref{int-as}) yields 
the following upper bound on the total number of atoms we may consider 
in our simulations 
\be
N\ll\lambda^{1/3}\sqrt{2\pi}\frac{l_z}{a_{\rm tt}},
\label{Nmax}
\ee
where $a_{\rm tt}$ is the $s$-wave scattering length for collisions between 
trapped atoms. 

\subsection{The atomic continuum}
The quadratic dependence of $\omega_k$ on $k$ is responsible for 
a number of mathematical difficulties arising in the context of atom 
lasers \cite{moypra299}. For the one-dimensional 
atom-laser model under consideration, the density of 
states which are available to a free atom diverges at frequencies 
close to $\omega_{\rm e}=0$, and is of the form 
\be
\rho(\omega)=\sqrt{\frac{m}{2\hbar\omega}}\Theta(\omega),
\label{dos}
\ee
where $\Theta(\omega)$ is the usual step function. Taking advantage of the 
symmetrical shape of the coupling and the even parity of $\omega_k$,
the spectral response of the continuum for the 
particular outcoupling mechanism under consideration reads
\be
D(\omega)\equiv 2|g(\omega,t)|^2\rho(\omega)=
\frac{\sqrt{2}\Lambda(t)}{\sqrt{\pi\omega_z}}
\frac{e^{-2\omega/\omega_z}}{\sqrt{\omega}}\Theta(\omega),
\label{srG}
\ee
where $g(\omega,t)$ is readily obtained from $g(k,t)$ using the atomic 
dispersion relation (\ref{disrel}).

Clearly, the spectral response (\ref{srG})  does not vary 
slowly for all the frequencies and thus we are dealing with a structured 
continuum which, in general, invalidates both Born and Markov approximations. 
To address fundamental mathematical difficulties associated 
with a continuum of this type, a number of new theoretical techniques 
have been developed \cite{bre99,str99,gara97,jack01} over 
the last years. In the following, 
we adopt a discretization approach first developed in the context 
of photonic band-gap continua \cite{nikolg}. 
In the case of atom-lasers, the same technique has been shown 
capable of providing 
not only the evolution of the number of atoms in the condensates, 
but also the distribution of the outcoupled atoms in frequency domain, 
irrespective of the strength of the outcoupling and the form of the 
spectral response \cite{NLP03}.

\section{Heisenberg equations of motion}
\label{SecIII}
Given the Hamiltonian (\ref{fullham}) one may proceed to derive 
Heisenberg equations of motion for the operators pertaining to the 
two traps and the continuum:
\numparts
\bea
\fl\frac{{\rm d}\aver{\hat{a}}}{{\rm d}t}=-{\rm i}\frac{\omega_z}2\aver{\hat{a}}
-2{\rm i}\kappa\aver{\hat{a}^\dag\hat{a}\hat{a}}-{\rm i}J\aver{\hat{b}}
-2{\rm i}\int_0^\infty {\rm d}k g(k,t) \aver{\hat{c}_k},
\label{em1a-pro}\\
\fl\frac{{\rm d}\aver{\hat{b}}}{{\rm d}t}=-{\rm i}\frac{\omega_z}2\aver{\hat{b}}
-2{\rm i}\kappa\aver{\hat{b}^\dag\hat{b}\hat{b}}-{\rm i}J\aver{\hat{a}}
-2{\rm i}e^{-\eta^2}\int_0^\infty  {\rm d}k g(k,t) \aver{\hat{c}_k},
\label{em1b-pro}\\
\fl\frac{{\rm d}\aver{\hat{c}_k}}{{\rm d}t}=-{\rm i}\omega_k \aver{\hat{c}_k}
-{\rm i} g(k,t)\aver{\hat{a}}
-{\rm i} g(k,t)e^{-\eta^2}\aver{\hat{b}}. \label{em1c-pro} 
\eea
\endnumparts
Due to the presence of interatomic collisions we have the 
appearance of third-order 
correlation functions on the right-hand side of equations 
(\ref{em1a-pro})-(\ref{em1c-pro}). 
Thus, this set of equations is not closed,  
while consideration of differential equations for the third-order 
correlation functions leads to the appearance of terms of even 
higher order and so on. In general, there are no exact remedies 
for such mathematical problems, but an approximate solution can be 
always obtained by decorrelating higher-order correlation functions into 
products of lower ones. 

In the present work we decorrelate the third-order correlation 
functions appearing on the right-hand side of equations 
(\ref{em1a-pro})-(\ref{em1c-pro}) as follows:
$\aver{\hat{a}^\dag\hat{a}\hat{a}}\approx
\aver{\hat{a}^\dag}\aver{\hat{a}}\aver{\hat{a}}$  
and $\aver{\hat{b}^\dag\hat{b}\hat{b}}\approx 
\aver{\hat{b}^\dag}\aver{\hat{b}}\aver{\hat{b}}$. 
We thus obtain the following closed set of coupled nonlinear differential 
equations 
\numparts
\bea
\fl\frac{{\rm d}\aver{\hat{a}}}{{\rm d}t}&=&-{\rm i}
\left (\frac{\omega_z}2+2\kappa|\aver{\hat{a}}|^2\right )\aver{\hat{a}}
-{\rm i}J\aver{\hat{b}}
-2{\rm i}\int_0^\infty {\rm d}k g(k,t) \aver{\hat{c}_k},
\label{em1a}\\
\fl\frac{{\rm d}\aver{\hat{b}}}{{\rm d}t}&=&-{\rm i}
\left (\frac{\omega_z}2+2\kappa|\aver{\hat{b}}|^2\right )\aver{\hat{b}}
-{\rm i}J\aver{\hat{a}}
-2{\rm i}e^{-\eta^2}\int_0^\infty  {\rm d}k g(k,t) \aver{\hat{c}_k},
\label{em1b}\\
\fl\frac{{\rm d}\aver{\hat{c}_k}}{{\rm d}t}&=&-{\rm i}\omega_k \aver{\hat{c}_k}
-{\rm i} g(k,t)\aver{\hat{a}}
-{\rm i} g(k,t)e^{-\eta^2}\aver{\hat{b}}. \label{em1c} 
\eea
\endnumparts
Although such a decorrelation is, in general, a crude approximation, 
it turns out to be equivalent to a higher-order decorrelation 
(mean-field approximation) of the form 
$\aver{\hat{a}^\dag\hat{a}\hat{a}}\approx
\aver{\hat{a}^\dag\hat{a}}\aver{\hat{a}}$   
and $\aver{\hat{b}^\dag\hat{b}\hat{b}}\approx 
\aver{\hat{b}^\dag \hat{b}}\aver{\hat{b}}$, when the two BECs 
are initially considered to be in coherent states. 

It is worth noting here that, in the context of the semiclassical 
(mean-field) theory, the initial coherence of the condensate(s) is preserved in time. 
On the contrary, quantum models which go beyond the mean-field approximation  
predict a collapse of the condensate wavefunction 
(e.g., see \cite{TwoModeValid,collapse}).  The time of collapse  
$t_{\rm c}\approx 1/(2\sqrt{N}\kappa)$ 
increases as we increase $N$ with fixed $\kappa N$, and thus the validity of the 
mean-field theory also improves. In any case, it is obvious that any experimental 
investigations pertaining to atom-laser outcoupling have to be performed 
for sufficiently short times (i.e., for $t\ll t_{\rm c}$) so that the coherence 
of the atom laser is guaranteed. 
In this context, mean-field theory has been shown capable of providing 
rather accurate and simple interpretation of the experimental observations 
(see, for instance, \cite{john07,profile,rmhc05, mf_vs_exp}).  

The presence of nonlinearities on the right-hand side of equations 
(\ref{em1a})-(\ref{em1c}) complicates somewhat their solution. 
Nevertheless, as long as the continuum 
under consideration is smooth (i.e., slowly varying) it can be eliminated 
by means of standard approximations (e.g., pole approximation). 
The atomic density of states (\ref{dos}) can be considered 
as smooth only for frequencies well above $\omega_{\rm e}$ and for 
sufficiently weak outcoupling rates. The latter assumption, however, 
is not always satisfied in typical experimental setups 
(e.g., see Ref. \cite{rmhc05,strong_exp}) and thus, 
the set of equations (\ref{em1a})-(\ref{em1c}) has to be treated exactly. 

To this end, we adopt here a uniform discretization approach. More precisely, 
we substitute the atomic continuum for frequencies within a range 
around $\omega_{z}/2$ (i.e., for $0<\omega<\omega_{\rm up}$), by a number 
(say $M$) of discrete modes, while the rest of the atom-mode density is 
treated perturbatively since 
it is far from resonance. The frequencies of the discrete modes are chosen to be 
$\omega_j=j\varepsilon$, where the mode spacing $\varepsilon$ 
is determined by the upper-limit condition of the discretization, namely
$\omega_{\rm up}=M\varepsilon$. The corresponding coupling for the $j$ mode, 
is determined by the spectral response (\ref{srG}) as follows
$\tilde{g}_j^2=D(\omega_j)\varepsilon$.
Discussion on the choice of $\omega_{\rm up}$ 
and the number of discrete modes can be found in Refs. \cite{nikolg}.
Working as in \cite{NLP03}, after the discretization equations 
(\ref{em1a})-\eref{em1c} read, 
\numparts
\bea
\fl\frac{{\rm d}\aver{\hat{a}}}{{\rm d}t}=
-{\rm i}\left [\mu_{\rm A}(t)/\hbar-S\right ]\aver{\hat{a}}
-{\rm i}(J-Se^{-\eta^2})\aver{\hat{b}}
-{\rm i}\sum_{j=1}^{M} \tilde{g}_j \aver{\hat{c}_j},
\label{em1a2}\\
\fl\frac{{\rm d}\aver{\hat{b}}}{{\rm d}t}=
-{\rm i}\left [\mu_{\rm B}(t)/\hbar-S e^{-2\eta^2}\right ]\aver{\hat{b}}
-{\rm i}(J-Se^{-\eta^2})\aver{\hat{a}}
-{\rm i}e^{-\eta^2}\sum_{j=1}^{M} \tilde{g}_j\aver{\hat{c}_j},
\label{em1b2}\\
\fl\frac{{\rm d}\aver{\hat{c}_j}}{{\rm d}t}=
-{\rm i}\omega_j \aver{\hat{c}_j}-{\rm i}\tilde{g}_j\aver{\hat{a}}
-{\rm i} \tilde{g}_j e^{-\eta^2} \aver{\hat{b}}, \label{em1c2} \eea
\endnumparts
with the shift given by 
$S=\int_{\omega_{\rm up}}^\infty {\rm d}\omega D(\omega)/\omega$, while we have 
introduced the chemical potentials for the two BECs namely,
$\mu_{\rm A}(t)=\hbar\omega_z/2+2\hbar \kappa |\aver{\hat{a}(t)}|^2$ and 
$\mu_{\rm B}(t)=\hbar\omega_z/2+2\hbar \kappa |\aver{\hat{b}(t)}|^2$. 
In the absence of 
interactions, the time-dependent nonlinearities 
vanish and we recover the equations of motion for the interaction-free 
model of  \cite{LazNikLam07} 
with $\mu_{\rm A}(t)=\mu_{\rm B}(t)=\hbar\omega_{z}/2$. 

\section{Simulations}
\label{SecIV}
Throughout our simulations we have considered $^{23}$Na BECs with 
$m=3.818\times 10^{-26}$ Kgr and $a_{\rm tt}=2.75\times 10^{-9}$ m, 
which are formed 
independently in identical harmonic traps with longitudinal oscillation 
frequency $\omega_z=200~\textrm{s}^{-1}$ and ratio $\lambda=0.4$. 
Condition (\ref{Nmax}) thus yields $N\ll 2.5\times 10^{3}$ i.e., 
our two-mode model is valid for relatively small BECs consisting 
of a few hundred of atoms. 
    
We assume that the BECs A and B are initially prepared in coherent 
states $\ket{\alpha}$ and $\ket{\beta}$, respectively.
Equations (\ref{em1a2})-(\ref{em1c2}) are thus solved with initial conditions 
\bea
&&\aver{\hat{a}(0)}=\alpha=\sqrt{N\tilde{\alpha}(0)},
\nonumber\\
&&\aver{\hat{b}(0)}=\beta=\sqrt{N\tilde{\beta}(0)} e^{{\rm i}\phi(0)}, 
\nonumber\\
&&\aver{\hat{c}_j(0)}=0,
\label{init}
\eea
where $\phi(0)$ is the initial relative phase between the two BECs. 
Accordingly, the initial number of condensed atoms in the traps A and B are   
given by $N_{\rm A}(0)=|\aver{\hat{a}(0)}|^2=N\tilde{\alpha}(0)$ and 
$N_{\rm B}(0)=|\aver{\hat{b}(0)}|^2=N\tilde{\beta}(0)$ respectively. 
Introducing the amplitude for the atomic continuum $\tilde{\gamma}$, 
at any time $t\geq 0$ we have 
$\tilde{\alpha}(t)+\tilde{\beta}(t)+\tilde{\gamma}(t)=1$, so that 
$N_{\rm trap}(t)+N_{\rm C}(t)=N$, where 
$N_{\rm trap}(t)\equiv N_{\rm A}(t)+N_{\rm B}(t)$
and $N_{\rm C}(t)$ denote the population of the traps and continuum, 
respectively.
Finally, for the sake of simplicity and without introducing any significant 
errors, the applied outcoupling pulse $\Lambda(t)$ is modeled as rectangular 
lasting from $t=0$ to $t=\tau$. From now on, the indices A(B) and C, refer to 
traps A(B) and the continuum, respectively. 

The dynamics of the system in the framework of an interaction-free model and for 
$\phi(0)=0$ were analyzed in detail elsewhere \cite{LazNikLam07}. 
For the sake of completeness, in the following 
subsection we recapitulate briefly the main results of \cite{LazNikLam07}. 

\subsection{Interaction-free model and vanishing relative phase}  
For the particular parameters under consideration, the weak-outcoupling 
regime corresponds to outcoupling rates $\Lambda<5\times 10^{2}~{\rm s}^{-2}$ 
\cite{LazNikLam07}. In this regime, the system evolves in time with the BECs 
exchanging population but the oscillations are exponentially 
damped as atoms are irreversibly coupled out of the traps.  The strength 
of the oscillations is determined by the Josephson coupling whereas the 
decay rate by the outcoupling strength $\Lambda$. Moreover, the 
distribution of the outcoupled atoms exhibits a characteristic asymmetric 
doublet with the two peaks separated by $2J$. 

The strong-outcoupling regime occurs for 
$\Lambda\gtrsim 5\times 10^{2}~{\rm s}^{-2}$  and is characterized by 
a continuous population exchange between the two BECs 
as well as between the BECs and the continuum. In the latter process, 
the pumping BEC  
seems to participate passively as its population is gradually transferred to 
the continuum in an irreversible way. 
On the contrary, the lasing BEC keeps exchanging population with the 
continuum even for 
larger times while for sufficiently strong outcoupling rates it is found only 
partially depleted in the long-time limit. The formation of such a bound state 
between the lasing condensate mode and the continuum is one of the most prominent 
effects associated with the non-Markovian nature of the dynamics.   
The distribution of the outcoupled atoms exhibits a broad peak, stemming from 
atoms directly outcoupled from the lasing BEC, while on the other hand 
atoms outcoupled from the pumping BEC give rise to a narrow peak.  
Most importantly, as a result of destructive 
quantum interference between the outcoupling channels in the system, 
the atomic distribution may also exhibit a dark spectral line (dip).  

We turn now to the discussion of the dynamics of the system in the case of 
a non-vanishing initial relative phase between the two BECs, as well as 
in the presence of interatomic interactions.

\subsection{Effects of relative phase}  
In order to understand the role of the initial relative phase $\phi(0)$, 
we solved equations (\ref{em1a2})-(\ref{em1c2}) numerically for increasing 
$\phi(0)\in [0,2\pi]$. For the sake of comparison with the 
results presented in \cite{LazNikLam07}, throughout our simulations  
we have focused on an interaction-free model. Moreover, we have considered 
various outcoupling rates $\Lambda$, in order to cover both 
Markovian and non-Markovian dynamics. 
Nevertheless, before we proceed to such complicated simulations, 
it is worth recalling here that in the absence of atomic collisions and 
outcoupling (i.e., for $\kappa=0$ and $\Lambda=0$), 
the evolution of the system is governed by the Josephson tunneling 
between the two BECs only. In this special case, the set of 
equations (\ref{em1a2})-(\ref{em1c2}) can be easily solved analytically, 
obtaining for the trapped populations
\bea
\fl\aver{a^\dag(t) a(t)}=N_{\rm A}(0)\cos^2(J t)+N_{\rm B}(0)\sin^2(J t)
+\sqrt{N_{\rm A}(0)N_{\rm B}(0)}\sin(2J t)\sin[\phi(0)]
\label{inter}
\eea
and $\aver{b^\dag(t) b(t)}=N-\aver{a^\dag(t) a(t)}$.
The first two terms in equation (\ref{inter}), describe the  
population exchange between the two traps when initially 
$N_{\rm A}(0)\neq N_{\rm B}(0)$, whereas for $N_{\rm A}(0)=N_{\rm B}(0)$ 
the time evolution of the system is governed only 
by the third term which is proportional to $\sin[\phi(0)]$. 
This term describes interference phenomena between the two BECs 
\cite{interef}
and was not present in the simulations performed in \cite{LazNikLam07}, 
as it vanishes for $\phi(0)=0$. In the following, we discuss 
the atom laser dynamics for various values of $\phi(0)$.

\subsubsection{Weak outcoupling---Markovian dynamics.} 
\begin{figure}
  \begin{center}
    \leavevmode
    \epsfxsize12.0cm
    \epsfbox{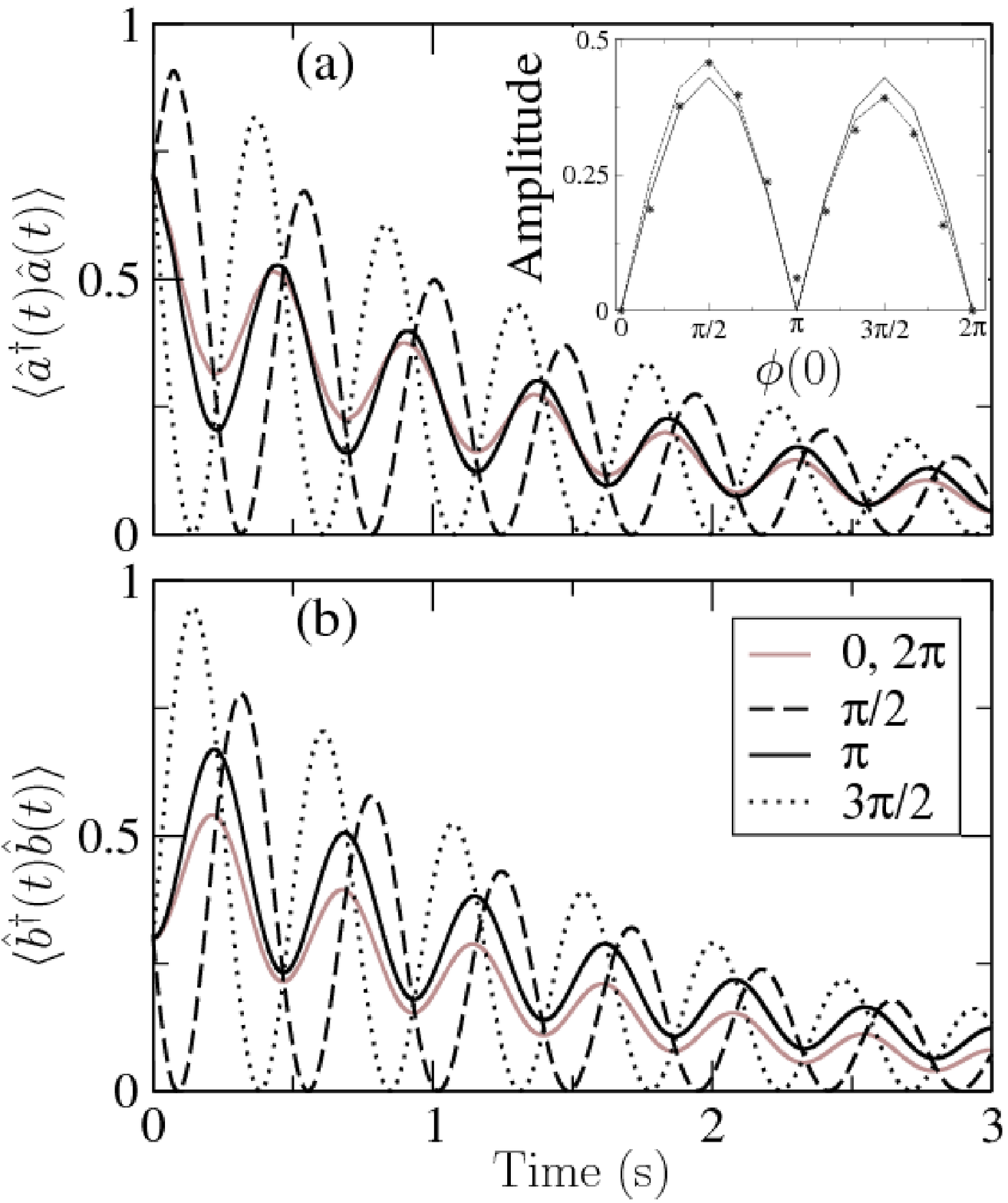}
  \end{center}
\caption{Effects of the relative phase on the Markovian dynamics. 
The normalized trapped populations are plotted as functions of time 
for various values of the initial relative phase $\phi(0)$ 
between the two BECs.  
Inset: Typical behavior of the amplitude of the oscillations 
for varying initial relative phase $\phi(0)$. 
The depicted values (stars) are estimated from the evolution 
of the trapped population $|\aver{\hat{a}^\dag(t) \hat{a}(t)}|^2$ 
for various $\phi(0)$ and in particular at its second peak. 
The corresponding amplitude for $\phi(0)=0$ is used as a reference 
amplitude. The solid and the dashed lines are sinusoidal 
fitting to the data, 
with constant $(\sim|\sin[\phi(0)]|)$  and exponentially damped amplitude 
$(\sim|\sin[\phi(0)]|e^{-0.05\phi(0)})$, respectively.  
Parameters: $\omega_z=200~{\rm s}^{-1}$, $\kappa=0$, 
$\lambda=0.4$, $\Lambda=10^2~{\rm s}^{-2}$, and $\eta=1.7$. 
Initial conditions: $\tilde{\alpha}(0)=0.7$,  $\tilde{\beta}(0)=0.3$. 
Discretization parameters: $M=1500$, $\omega_{\rm up}=300~{\rm s}^{-1}$.
}
\label{MPphase:fig}
\end{figure}

In the weak-outcoupling regime, the damped oscillatory behavior of the trapped 
populations reflects the Josephson tunneling between the two BECs, and the  
irreversible outcoupling of atoms into the continuum. As shown 
in \cite{LazNikLam07}, the evolution of the system under both of these 
physical processes strongly depends on the trap separation as well 
as the frequencies and the outcoupling rates of the two condensate modes.   
In addition, the present simulations suggest a strong dependence 
on the initial relative phase between the two BECs, which affects 
mostly the exchange of atoms between them. 
As depicted in figure \ref{MPphase:fig}, the amplitude of the oscillations  
varies with $\phi(0)$, while in the case of $\phi(0)=\pi/2,~3\pi/2$ 
and for short times, a large fraction of the total population (more than  80\%) may be 
found to occupy one of the traps only [see dashed and dotted 
lines in figures \ref{MPphase:fig}(a) and (b), respectively]. 

According to the inset of figure \ref{MPphase:fig}(a), 
the amplitude of the oscillations varies 
sinusoidally with $\phi(0)$; a behavior which can be explained, to some extent, 
in terms of the interference term 
appearing in equation (\ref{inter}), which is also proportional to $\sin[\phi(0)]$. 
Of course  equation (\ref{inter}) has been derived in the framework of 
a particularly simple model pertaining to two isolated BECs, and as such 
cannot describe all aspects of the observed dynamics. 
For instance, it is obvious that the amplitude variation depicted in 
the inset of figure \ref{MPphase:fig}(a) is not 
symmetric around $\phi(0)=\pi$ while even the solutions we have obtained 
for $\phi(0)=0$ and $\phi(0)=\pi$ do not coincide, 
despite the fact that the interference term in  equation (\ref{inter}) vanishes in  
both cases. Such deviations from the perfect sinusoidal behavior 
can be attributed to the presence of the outcoupling mechanism 
which is not included in the simple model that equation (\ref{inter}) refers to. 
More precisely, in the present scenario the two condensate modes are not isolated, 
but rather coupled to the same continuum. As a result, we have various outcoupling 
channels for the system which may also interfere, 
thus giving rise to phenomena that cannot be described 
in the context of isolated BECs.

\begin{figure}
  \begin{center}
    \leavevmode
    \epsfxsize12.0cm
    \epsfbox{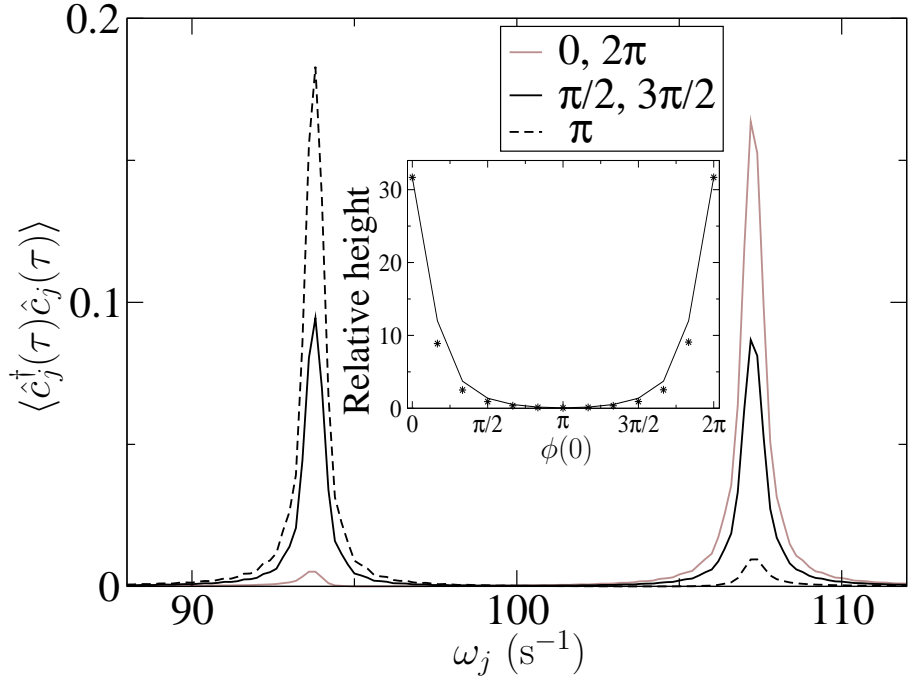}
  \end{center}
\caption{Effects of the relative phase on the distribution of outcoupled 
atoms in the Markovian regime. The distributions have been 
obtained at  $\tau=10~{\rm s}$ for various values of the 
initial relative phase $\phi(0)$ between the two BECs. 
Inset: Typical behavior of the relative height of the two peaks 
(ratio of the right to the left) for varying initial relative phase $\phi(0)$. 
The depicted values (stars) are estimated from the atomic distributions 
at $\tau=10~{\rm s}$ for various $\phi(0)$. 
The solid line is a fitting to the data according to ratio (\ref{rat}).
Parameters as in figure \ref{MPphase:fig}.}
\label{MSphase:fig}
\end{figure}

The distribution of the outcoupled atoms also varies with $\phi(0)$.  
More precisely, the characteristic doublet structure persists  
for any choice of $\phi(0)$, but the relative height of the peaks 
(ratio of the right to the left peak) changes periodically.  
For instance, in figure \ref{MSphase:fig} we see that 
for $\phi(0)=0$ (gray line), the right peak is more pronounced than the left peak, 
whereas the situation is the opposite for $\phi(0)=\pi$ (dashed line). Moreover, 
for $\phi(0)=\pi/2,~3\pi/2$ the two peaks have nearly the same heights 
(solid black line). 
For a better overview of this periodic change, 
in the inset of figure \ref{MSphase:fig}  we present the relative height of the 
two peaks for various values of $\phi(0)$ (stars). The observed behavior can 
be easily understood in terms of the global symmetric and antisymmetric states 
$\vert\pm\rangle$ of the double-well potential \cite{LazNikLam07}. 
The initial populations of the states are  
$P_{\pm}=\vert \sqrt{\tilde{\alpha}(0)} \pm 
\sqrt{\tilde{\beta}(0)}e^{{\rm i}\phi(0)}\vert^{2}/2$
and thus their normalized ratio reads 
\be
\frac{P_{+}}{P_{-}}= \frac{1+2\sqrt{\tilde{\alpha}(0)\tilde{\beta}(0)}\cos[\phi(0)]}{1-2\sqrt{\tilde{\alpha}(0)\tilde{\beta}(0)}\cos[\phi(0)]}.
\label{rat}
\ee
This expression describes the variation of the distribution 
of the initial population between the two global states with respect to 
$\phi(0)$.  As depicted in the inset of figure \ref{MSphase:fig}, 
the ratio (\ref{rat}) (multiplied by a constant) is in 
a very good agreement with  the periodic changes observed 
in the relative height of the two peaks in the distribution of the 
outcoupled atoms.

\subsubsection{Strong outcoupling---Non-Markovian dynamics.} 
\label{SNM-phase}
Interference effects between the two BECs may also alter the evolution of the 
system in the strong outcoupling regime. More precisely, by varying the 
initial relative phase between the two BECs we find that the most 
noticeable changes in the evolution of the trapped populations occur for 
relatively short times, as well as in the long-time limit.

\begin{figure}
  \begin{center}
    \leavevmode
    \epsfxsize12.0cm
    \epsfbox{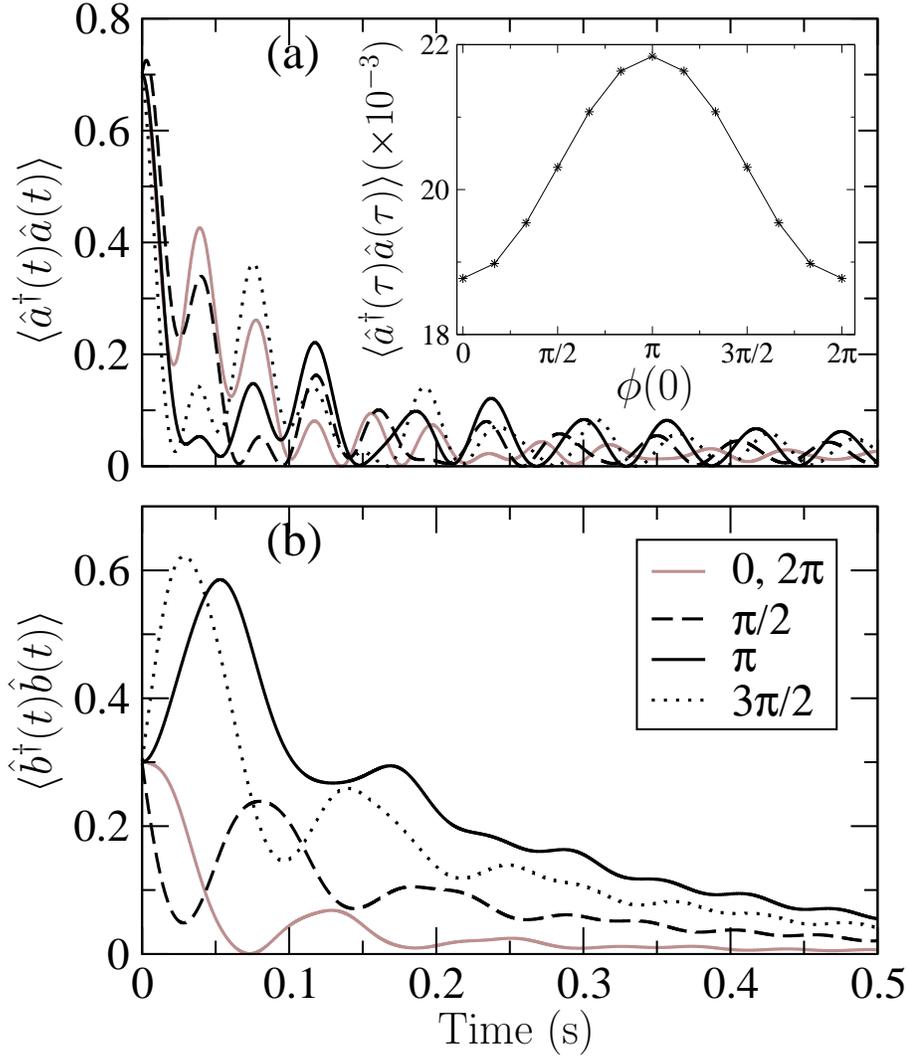}
  \end{center}
\caption{Effects of the relative phase on the non-Markovian dynamics. 
The normalized trapped populations are plotted as functions of time 
for various values of the initial relative phase $\phi(0)$ 
between the two BECs.  
Inset: Typical behavior of the steady-state population 
of the lasing condensate mode for varying initial relative phase between the two BECs. 
The depicted values (stars)  are estimated numerically at 
$\tau=10~{\rm s}$, whereas the solid line is a fitting of the 
form $x+y\cos[\phi(0)]$.
Parameters: $\omega_z=200~{\rm s}^{-1}$, $\kappa=0$, 
$\lambda=0.4$, $\Lambda=4\times 10^3~{\rm s}^{-2}$, and $\eta=1.5$. 
Initial conditions: $\tilde{\alpha}(0)=0.7$, $\tilde{\beta}(0)=0.3$. 
Discretization parameters: $M=1500$, $\omega_{\rm up}=300~{\rm s}^{-1}$.
}
\label{NMPphase:fig}
\end{figure}

Let us start with trap B for which we can identify two distinct regimes in 
the evolution of the corresponding population for $\phi(0)=0$ [see gray line in 
figure \ref{NMPphase:fig}(b)]. 
For short times the main part of the population is transferred into the continuum. 
After this initial transient regime, dissipation is temporarily 
turned off and a weak oscillatory population exchange between the two traps 
sets in. Nevertheless, trap B is empty in the long-time limit as 
its population is gradually transferred into the continuum in an irreversible, 
almost exponential, way. As depicted in figure \ref{NMPphase:fig}(b), the initial 
transient regime gradually disappears with increasing $\phi(0)$ and for 
$\phi(0)>5\pi/6$, it is replaced by population exchange between the two BECs.
In particular, for very short times the pumping BEC gets population from the 
lasing BEC, and this transfer becomes more pronounced as we increase $\phi(0)$ up to 
$\sim 4\pi/3$. After this initial population exchange, the condensate 
mode B decays almost exponentially and is empty in the long-time limit.  
Of course, increasing further $\phi(0)\in(4\pi/3,2\pi]$, we find that the initial 
transient regime is gradually resettled.

The situation is similar but more interesting in the case of trap A. 
More precisely, as depicted in figure \ref{NMPphase:fig}(a), the initial decay of 
the lasing condensate mode 
becomes faster and stronger as we increase $\phi(0)$ up to $\pi$. 
For $\phi(0)=0$, the main part of the population that 
leaves the lasing BEC in this initial transient regime is lost 
into the continuum. 
The situation, however, is different for $\phi(0)\neq 0$ since  
only part of the population is outcoupled while the remaining part 
is transferred to trap B. This transfer becomes more pronounced for 
$\phi(0)> 5\pi/6$ [see black solid and dotted lines in  
figures \ref{NMPphase:fig}(a,b)].  
In any case, after this initial transient regime, dissipation is temporarily 
turned off and the condensate mode A exchanges population with the 
continuum only. 
This oscillatory population exchange persists even for longer times and the 
amplitude of the oscillations varies with $\phi(0)$. In the long-time limit, we 
have the formation of a bound state between the lasing condensate mode and 
the continuum. As a result, trap A exhibits a non-zero 
steady-state population which, as depicted in the inset of figure 
\ref{NMPphase:fig}(a), also varies periodically with $\phi(0)$, 
and attains its maximum at $\phi(0)=\pi$. 
This behavior can be interpreted in terms of the global states 
$\ket{\pm}$ with energies $\omega_z/2\pm J$. 
For $\eta<2.0$, the Josephson coupling $J>0$ and thus, the antisymmetric 
state $\ket{-}$ is closer to the edge than the symmetric state. 
As a result, the antisymmetric state is more protected against 
dissipation and contributes the most to the steady-state population. 
Moreover, as we discussed earlier, the initial population of the antisymmetric 
state is given by $P_{-}= 0.5-\sqrt{\tilde{\alpha}(0)
\tilde{\beta}(0)}\cos[\phi(0)]$ and attains its maximum 
at $\phi(0)=\pi$. It is this periodic dependence of $P_{-}$ on $\phi(0)$, 
that is also reflected in the steady-state population of trap A 
[see inset of figure \ref{NMPphase:fig}(a)].

\begin{figure}
  \begin{center}
    \leavevmode
    \epsfxsize12.0cm
    \epsfbox{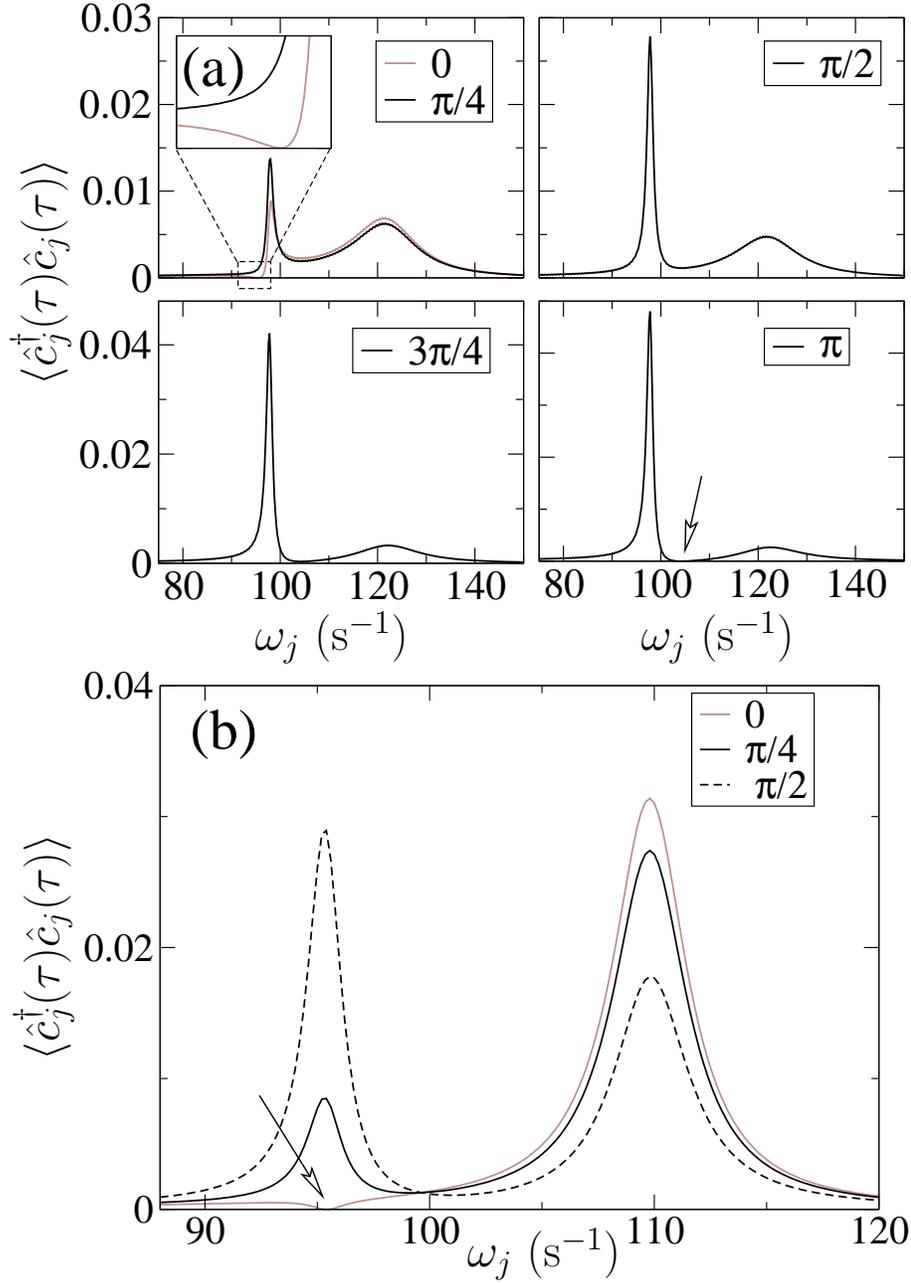}
  \end{center}
\caption{Effects of the relative phase on the distribution of outcoupled atoms 
in the non-Markovian regime. The distributions have been obtained at 
$\tau=10~{\rm s}$ for various initial values of the relative phase $\phi(0)$ 
between 
the two BECs. Other parameters as in figure \ref{NMPphase:fig}, 
but for $\eta=1.7$  and $\Lambda=2\times 10^{3}~{\rm s}^{-2}$ (a); 
$\eta=1.7$  and $\Lambda=5\times 10^{2}~{\rm s}^{-2}$ (b). The arrows point at 
the dips while the inset shows a closeup of the atomic distribution around it.
}
\label{NMSp:fig}
\end{figure}

Finally, the choice of the initial relative phase may also affect significantly 
certain aspects of the distribution of outcoupled atoms in the strong-outcoupling 
regime. 
More precisely, the first aspect pertains to the relative height of the 
two peaks. 
As discussed in  \cite{LazNikLam07}, the narrow peak stems from atoms directly 
outcoupled from the pumping BEC, 
whereas the broad peak is associated with atoms that emerge from the 
lasing BEC and as such experience stronger outcoupling. Thus, any variations 
in the population transfer between the two BECs with respect $\phi(0)$,  
are expected to be reflected in the relative height of the two peaks. 
Indeed, as discussed above, the fraction of the atoms 
transferred from the lasing to the pumping mode 
increases as we increase $\phi(0)\in [0,\pi]$.  
Accordingly, in this regime of relative phases the narrow peak related to the 
exponential decay of trap B becomes higher and more pronounced for increasing 
$\phi(0)$ at the detriment of the broad peak. 
Increasing $\phi(0)$ further, the narrow peak decreases again whereas 
the broad peak becomes stronger.
The second aspect of the atomic distribution which is also of particular 
interest, involves the presence of a dark line (dip). As depicted in the inset 
of figure \ref{NMSp:fig}(a), the 
dip is present only $\phi(0)=0~{\rm or}~2\pi$, while a new dip appears between 
the two peaks for $\phi(0)=\pi$.  These observations support  
the explanation for the dip given in \cite{LazNikLam07} namely, 
it is a signature of destructive interference between the various outcoupling 
channels in the system. The degree as well as the nature of the interference 
is expected to vary with the relative phase between the two BECs. 
For instance, when $\Lambda=5\times10^2{\rm s}^{-2}$, 
the dip appears to give its place to a peak which is now a 
signature of constructive interference between various outcoupling channels 
[see figure \ref{NMSp:fig}(b)].

\subsection{Effects of interatomic interactions}
In general, the evolution of the system under consideration is governed by 
three distinct physical processes. More precisely, apart from the  
Josephson and the output coupling which were also present in the 
interaction-free model of \cite{LazNikLam07}, we also have the repulsive collisional 
interactions. It is reasonable therefore to define the ratios 
$\nu=\kappa N/J$ and $\xi=\kappa N/\sqrt{\Lambda}$ 
which quantify the effect of interatomic interactions relative
to tunneling and outcoupling effects, respectively. 
How strongly the inclusion of interactions affects the 
results obtained in the framework of the interaction-free model of 
\cite{LazNikLam07} depends on these two ratios. In particular,  
the largest deviations are expected for $\nu\gtrsim 1$ and $\xi \gtrsim 1$.

Before proceeding with our simulations, it is worth recalling here 
the dynamics of two isolated tunnel-coupled BECs 
[as determined by equations (\ref{em1a2})-(\ref{em1c2}) for $\Lambda=0$]. 
In this case, the trapped population 
is conserved i.e., $N_{\rm trap}(t)=N$, while for a given initial population 
imbalance in the system $\tilde{p}(0)\equiv (N_{\rm A}(0)-N_{\rm B}(0))/N$, 
we have complete periodic oscillations between the two condensates 
({\em Josephson regime}) \cite{java86} when \cite{ragh99}   
\be
H_{\rm c}(0)\equiv \frac{\nu}{2}\tilde{p}(0)^2+\sqrt{1-\tilde{p}(0)^2}
\cos[\phi(0)]<1. 
\label{nci}
\ee 
The period of the oscillations increases with increasing nonlinearity 
(i.e., with increasing $N$) while the oscillations become anharmonic  
as $H_{\rm c}(0)\to 1$. Finally, for $H_{\rm c}(0)>1$ the oscillations 
are no longer complete while their period reduces with increasing 
nonlinearity ({\em self-trapping regime}) \cite{SelfTrapping}. 

One can readily check that $H_{\rm c}(t)$ is a constant of motion in the 
absence of outcoupling. 
In our system, however, the situation is more involved as atoms are 
outcoupled from the two BECs with different rates and thus neither 
$N_{\rm trap}$ nor $H_{\rm c}$ is conserved.  
In the following, we discuss separately the dynamics of the 
system for weak and strong outcoupling rates when it starts from 
the Josephson and the self-trapping regime. As we will see, during its 
evolution the system may pass from one regime to the other. 
 
\subsubsection{Weak outcoupling---Markovian dynamics.} 
In this subsection we investigate how the inclusion of interactions in our model 
affects the dynamics of the system in the weak-outcoupling regime. 
For the sake of comparison with the interaction-free model discussed 
in \cite{LazNikLam07}, we will focus on the case of vanishing initial 
relative phase between the two BECs.   

\begin{figure}
  \begin{center}
    \leavevmode
    \epsfxsize12.0cm
    \epsfbox{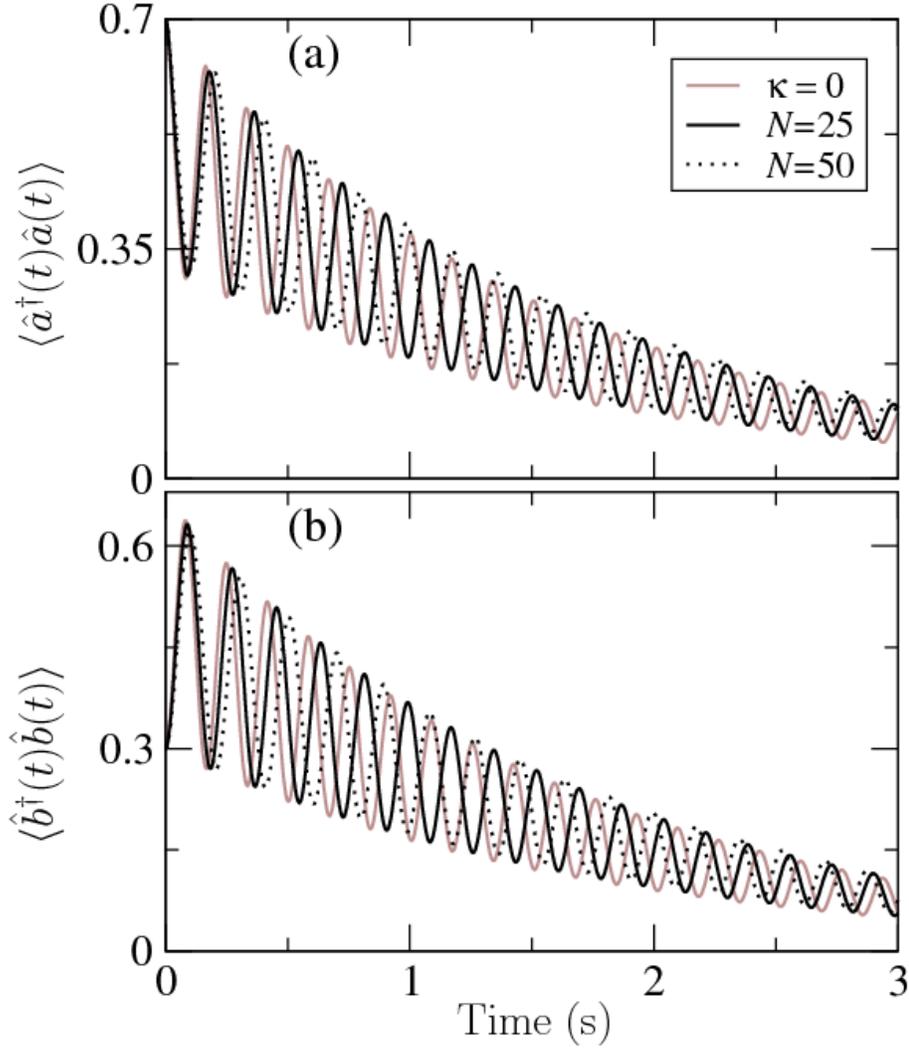}
  \end{center}
\caption{
Effects of interatomic interactions on the Markovian dynamics. 
The normalized trap populations are plotted as functions of time 
for various atom numbers and $\omega_z=200~{\rm s}^{-1}$, 
$\lambda=0.4$, $\Lambda=10^2~{\rm s}^{-2}$, and $\eta=1.5$. 
The gray curves correspond to an interaction-free Bose gas. 
Initial conditions: Josephson regime with 
$\tilde{\alpha}(0)=0.7$,  $\tilde{\beta}(0)=0.3$, $\phi(0)=0$. 
Discretization parameters: $M=1500$, $\omega_{\rm up}=300~{\rm s}^{-1}$.}
\label{markP:fig}
\end{figure}

\begin{figure}
  \begin{center}
    \leavevmode
    \epsfxsize12.0cm
    \epsfbox{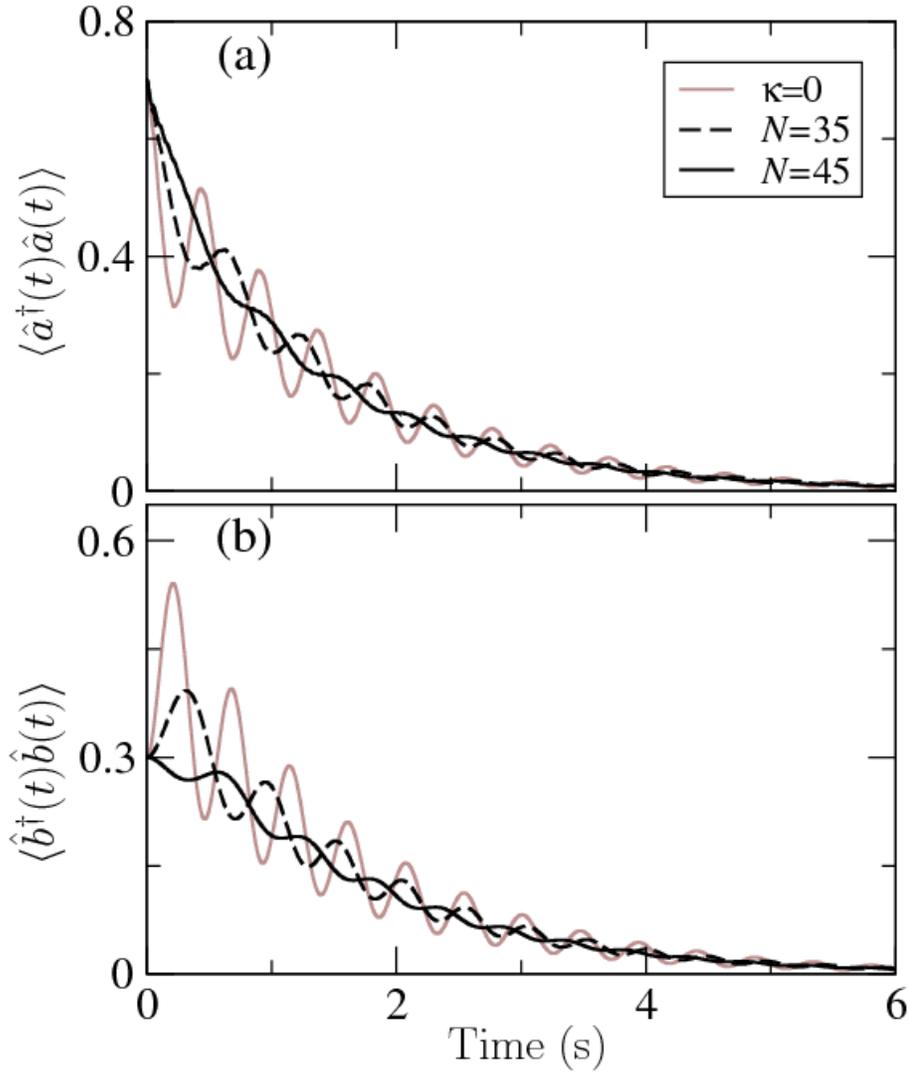}
  \end{center}
\caption{ As in figure \ref{markP:fig} for $\eta=1.7$.} \label{markP2:fig}
\end{figure}

When the system starts from the Josephson regime [i.e., if $H_{\rm c}(0)<1$] 
and $\nu<1$, the evolution of the trapped populations in the presence 
of interactions is qualitatively similar to the evolution we obtained 
in the context of the interaction-free model. 
The small quantitative differences depicted in figure \ref{markP:fig}  
are due to interatomic interactions which enter the 
equations of motion for $\aver{\hat{a}(t)}$ and $\aver{\hat{b}(t)}$ as 
time-dependent nonlinearities proportional to the corresponding 
trapped populations $N_{\rm A}(t)=|\aver{\hat{a}(t)}|^2$ and 
$N_{\rm B}(t) =|\aver{\hat{b}(t)}|^2$ 
[see equations (\ref{em1a2}) and (\ref{em1b2})]. 
As a result, the chemical potentials of the two BECs fluctuate   
in time and become off-resonant. As long as $\nu<1$  
the nonlinearities are so weak that may only shift the oscillations in the 
populations with respect to the oscillations in the case of the ideal 
gas, whereas for $\nu\gtrsim 1$  the nonlinearities 
are sufficiently strong to disturb the exchange of population 
between the two BECs. In particular, as 
is evident in figure \ref{markP2:fig}, the oscillatory behavior 
of both occupation probabilities becomes less pronounced as we 
increase the  nonlinearity in the system.  Thus for sufficiently 
strong nonlinearities the two traps tend to decay nearly 
exponentially into the atomic continuum 
(see solid curves in figure \ref{markP2:fig}).

\begin{figure}
  \begin{center}
    \leavevmode
    \epsfxsize12.0cm
    \epsfbox{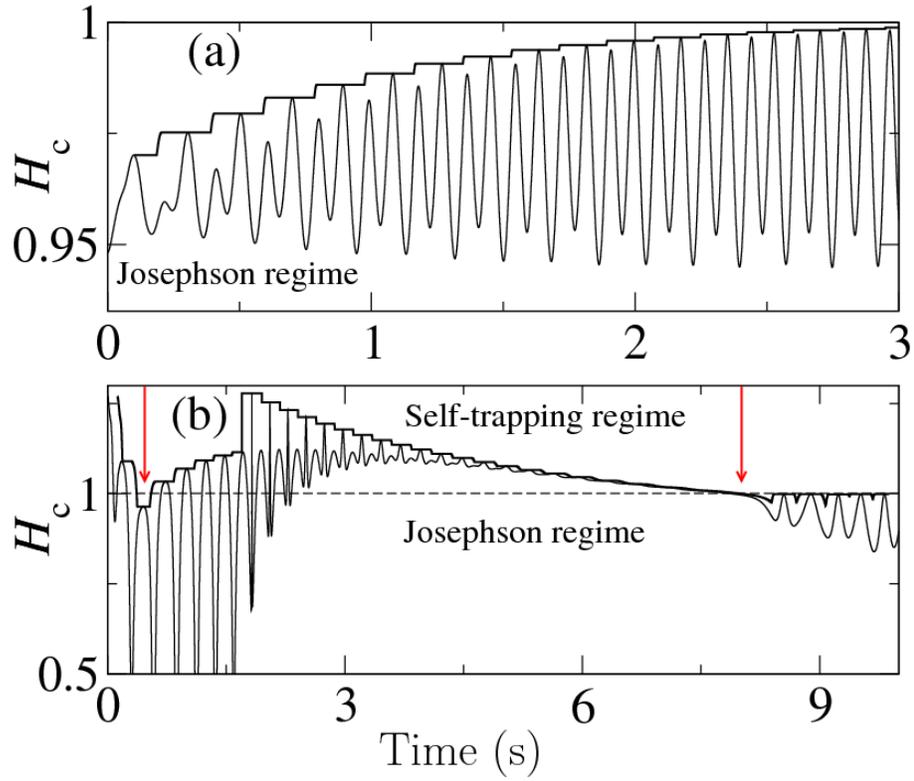}
  \end{center}
\caption{The evolution of $H_{\rm c}$ (thin line) 
and its envelope (thick line), are plotted as functions of time 
for the parameters of figure \ref{markP:fig}(a) ($N=50$) 
and figure \ref{markP3:fig}(d).  
The arrows point at the regions where the system passes from one regime 
to the other.} 
\label{mst:fig}
\end{figure}

The question that arises here is whether the system remains in the Josephson 
regime throughout its evolution or not. To answer this question, 
one may follow, for instance, the time evolution of 
$H_{\rm c}(t)$ [with $\nu=\kappa N_{\rm trap}(t)/J$, 
$\tilde{p}(t)\equiv (N_{\rm A}(t)-N_{\rm B}(t))/N$, 
$\phi(t)=\arg(\aver{a^\dag b})$] which is not a conserved 
quantity anymore, as atoms are continuously coupled out of the traps. 
In figure \ref{mst:fig} we plot the evolution of $H_{\rm c}(t)$ 
as a function of time, for the parameters of figure \ref{markP:fig}(a) 
(with $N=50$).  
Clearly, we have $H_{\rm c}(t)<1$ for all $t$, and thus we may safely 
conclude that the system remains in the Josephson regime throughout its 
evolution. The same conclusion can be drawn by following the evolution of 
$H_{\rm c}(t)$ for the parameters of the other plots in figures 
\ref{markP:fig} and \ref{markP2:fig}

\begin{figure}
  \begin{center}
    \leavevmode
    \epsfxsize12.0cm
    \epsfbox{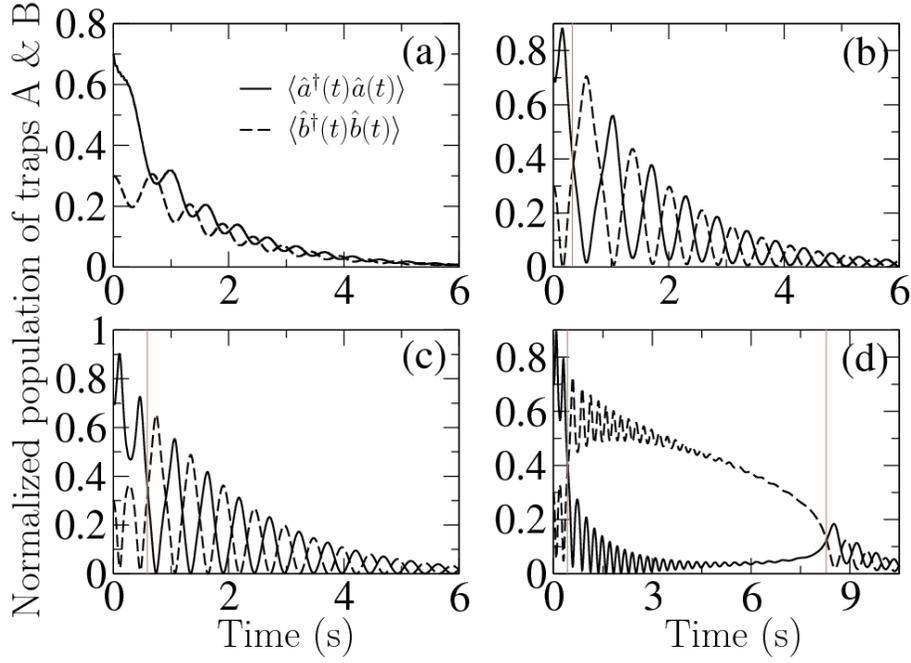}
  \end{center}
\caption{
Effects of interatomic interactions on the Markovian dynamics. 
The normalized trap populations are plotted as functions of time 
for various atom numbers: (a) $N=50$; (b) $N=100$; (c) $N=150$; (d) $N=200$. 
The vertical gray line denotes the border between self-trapping 
and Josephson regime.  
Parameters: $\omega_z=200~{\rm s}^{-1}$, $\lambda=0.4$, 
$\Lambda=10^2~{\rm s}^{-2}$, $\eta=1.7$.
Initial conditions:  self-trapping 
regime with $\tilde{\alpha}(0)=0.7$,  $\tilde{\beta}(0)=0.3$, $\phi(0)=0$. 
Discretization parameters: $M=1500$, $\omega_{\rm up}=300~{\rm s}^{-1}$.
} \label{markP3:fig}
\end{figure}

Let us consider now the situation where the system starts from the 
self-trapping regime [i.e., $H_{\rm c}(0)>1$] . 
As depicted in figure  \ref{markP3:fig}, for short-times 
the occupation probabilities for the two traps undergo partial oscillations 
which typically characterize the self-trapping regime. Gradually, however, 
the population imbalance between the two traps reduces as atoms are 
coupled out of trap A, $e^{\eta^2}$ times faster than trap B and after some 
time we have $N_{\rm A}=N_{\rm B}$ (vertical gray lines). In the absence of 
losses 
(i.e., for $\Lambda=0$) such a balanced state can be a stationary 
state of the nonlinear Schr\"odinger equation. In the present case, however, 
the system is driven again away from this state as the outcoupling mechanism 
is still 
on and the population balance is immediately destroyed. For sufficiently weak 
nonlinearities, the balanced state marks the transition from the 
self-trapping to the Josephson regime where we have complete damped oscillations 
between the two traps [e.g., see figures \ref{markP3:fig}(b,c)]. 
On the contrary, for larger nonlinearities [see figure \ref{markP3:fig}(d)], 
the system tends to continue  
in the self-trapping regime up to the point the population imbalance between 
the two traps vanishes again. 
These transitions are also evident in the evolution of the corresponding 
$H_{\rm c}(t)$, 
presented in figure \ref{mst:fig}(b). In contrast to figure \ref{mst:fig}(a), 
we see here that $H_{\rm c}(t)$ exhibits abrupt and instantaneous changes,  
which of course are not reflected in the evolution of the corresponding 
populations. 
However, the profile of $H_{\rm c}(t)$ (thick line) characterizes rather 
accurately the transition points discussed above. 

According to our simulations, the ratio $\xi$ does not seem to affect 
considerably the oscillatory behavior of the occupation probabilities 
for the two traps. This was to be expected to some extent, as we are in the 
weak-outcoupling regime and thus these oscillations reflect the 
population exchange between the two BECs. Hence, it is the ratio 
$\nu$ the one that determines how strongly the population exchange is 
affected by the presence of interactions in the system. 
Nevertheless, our simulations show that the distribution of the outcoupled 
atoms is determined by the ratio $\xi$. 

\begin{figure}
  \begin{center}
    \leavevmode
    \epsfxsize12.0cm
    \epsfbox{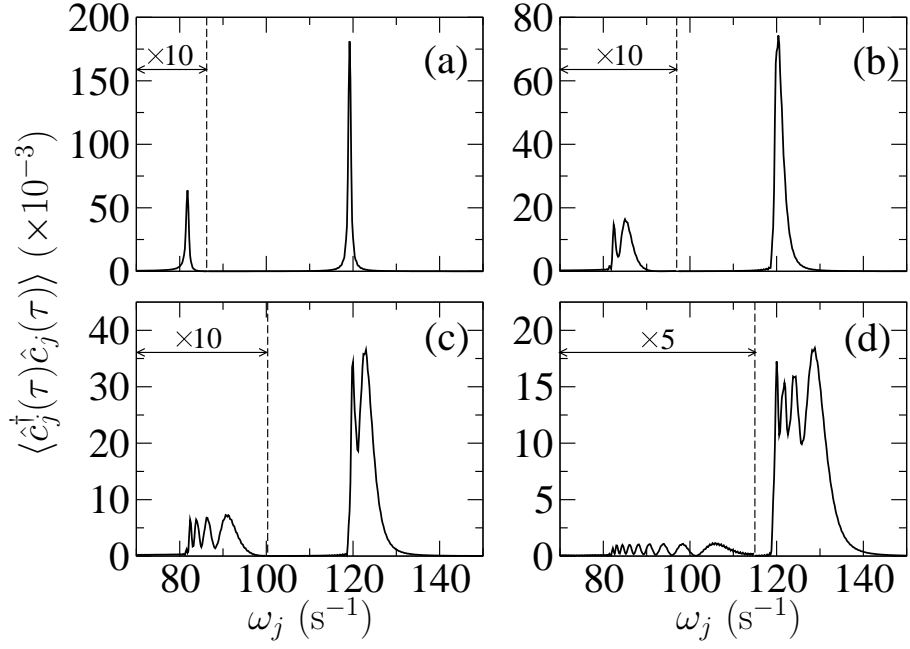}
  \end{center}
\caption{
Effects of interatomic interactions on the distribution of outcoupled atoms 
in the Markovian regime. 
The distribution (a) corresponds to an interaction-free Bose gas 
whereas the other three distributions have been obtained for a 
weakly interacting Bose gas with $N=25$ (b),  $N=50$ (c), $N=100$ (d).
Parameters: $\omega_z=200~{\rm s}^{-1}$, $\lambda=0.4$, 
$\Lambda=10^2~{\rm s}^{-2}$, $\eta=1.5$, $\tau=10~{\rm s}$. 
Initial conditions:  Josephson regime with 
$\tilde{\alpha}(0)=0.7$,  $\tilde{\beta}(0)=0.3$, $\phi(0)=0$. 
Discretization 
parameters: $M=1500$, $\omega_{\rm up}=300~{\rm s}^{-1}$.
} \label{MarkSpec:fig}
\end{figure}

In figure \ref{MarkSpec:fig} we depict the distribution of outcoupled atoms at 
$\tau=10~{\rm s}$ for various atom numbers, and let us focus for a while 
on the upper peak of the doublet. As we increase the number 
of atoms in the system, the peak becomes broader while its shape starts 
deviating from the standard Lorentzian profile. Actually, 
for $\xi\gtrsim 1$ it acquires a rather complicated structure. 
This behavior can be understood in terms of the time-dependent 
chemical potentials appearing in equations (\ref{em1a2}) and (\ref{em1b2}). 
More precisely, due to the applied outcoupling mechanism, the trapped 
population reduces with time. For very short times, the frequency of the 
outcoupled atoms that contribute to the upper peak is roughly $\mu_{\rm A}(0)/\hbar+J$ 
while as time goes on, $\mu_{\rm A}(t)$ changes due to the applied outcoupling mechanism 
as well as the coupling to the pumping condensate. 
Hence, outcoupled atoms may involve different frequencies and 
the center of the atomic distribution is shifted accordingly  towards lower or upper 
frequencies (``chirping''); a phenomenon which is reflected in the broadening of 
the upper peak.  
Moreover, interference effects between atoms which have the same frequency  
and were outcoupled at different times, may also give rise to unconventional 
atomic distributions exhibiting multiple spikes 
[see figures \ref{MarkSpec:fig}(c-d)]. 

Increasing the total number of atoms in the system $N$ or reducing the 
outcoupling rate $\sqrt{\Lambda}$, we essentially increase the ratio $\xi$ 
and thus the effects of interactions on the atom-laser outcoupling are 
more pronounced. In this spirit one may notice, for example, that the same 
interaction effects we have just described also appear on the lower peak, 
albeit sooner. The reason is simply that this peak is associated with 
atoms outcoupled directly from trap B with outcoupling rate 
$\sqrt{\Lambda}e^{-\eta^2}$. 

In closing, we would like to point out that similar conclusions can be 
drawn by following the evolution of the atomic distribution with time,  
for a fixed initial nonlinearity i.e., for a given $N$. In this case, 
the effects of 
interactions become more prominent as time goes on, and the distribution 
of the outcoupled atoms begins exhibiting sharp spikes due to quantum interference. 
Moreover, all atomic distributions we have checked throughout our simulations 
show the characteristics discussed above, irrespective of the chosen initial 
conditions. Finally, it is worth noting that the ``chirping'' effect and the 
unconventional atom laser spectra have also been 
predicted recently by Johnsson {\em et al.} \cite{john07},    
and a chirp-compensation mechanism has been proposed. 
Moreover, chirping effects may also occur in more realistic  
three-dimensional systems, disturbing the transverse density distribution 
of the atom-laser beam \cite{profile}.

\subsubsection{Strong outcoupling---Non-Markovian dynamics.} 
We turn now to the discussion of how the inclusion of interactions in 
our model affects 
the evolution of the system and the distribution of outcoupled atoms 
in the strong-outcoupling regime. As in the weak-outcoupling regime, 
we will focus on $\phi(0)=0$.

\begin{figure}
  \begin{center}
    \leavevmode
    \epsfxsize12.0cm
    \epsfbox{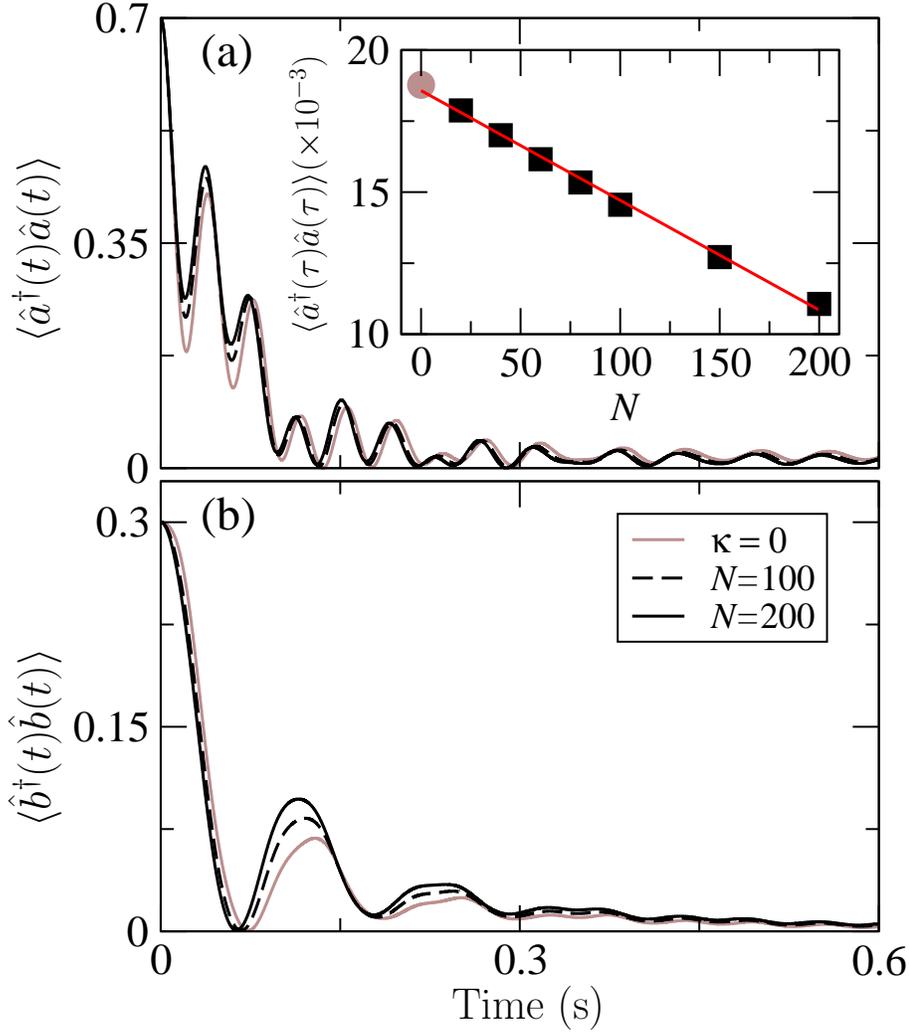}
  \end{center}
\caption{Effects of interatomic interactions on the non-Markovian dynamics. 
The normalized trap populations are plotted as functions of time 
for $\omega_z=200~{\rm s}^{-1}$, $\lambda=0.4$, 
$\Lambda=4\times 10^3~{\rm s}^{-2}$, $\eta=1.5$. 
The inset shows the typical behavior of the steady-state population 
of trap A at $\tau=10~{\rm s}$ for various atom numbers. 
The gray curves and circles correspond to an interaction-free Bose gas. 
Initial conditions: Josephson regime ($N=100$) and 
self-trapping regime ($N=200$) with 
$\tilde{\alpha}(0)=0.7$,  $\tilde{\beta}(0)=0.3$, $\phi(0)=0$. 
Discretization 
parameters: $M=1500$, $\omega_{\rm up}=300~{\rm s}^{-1}$.
} \label{nmarkP1:fig}
\end{figure}

\begin{figure}
  \begin{center}
    \leavevmode
    \epsfxsize12.0cm
    \epsfbox{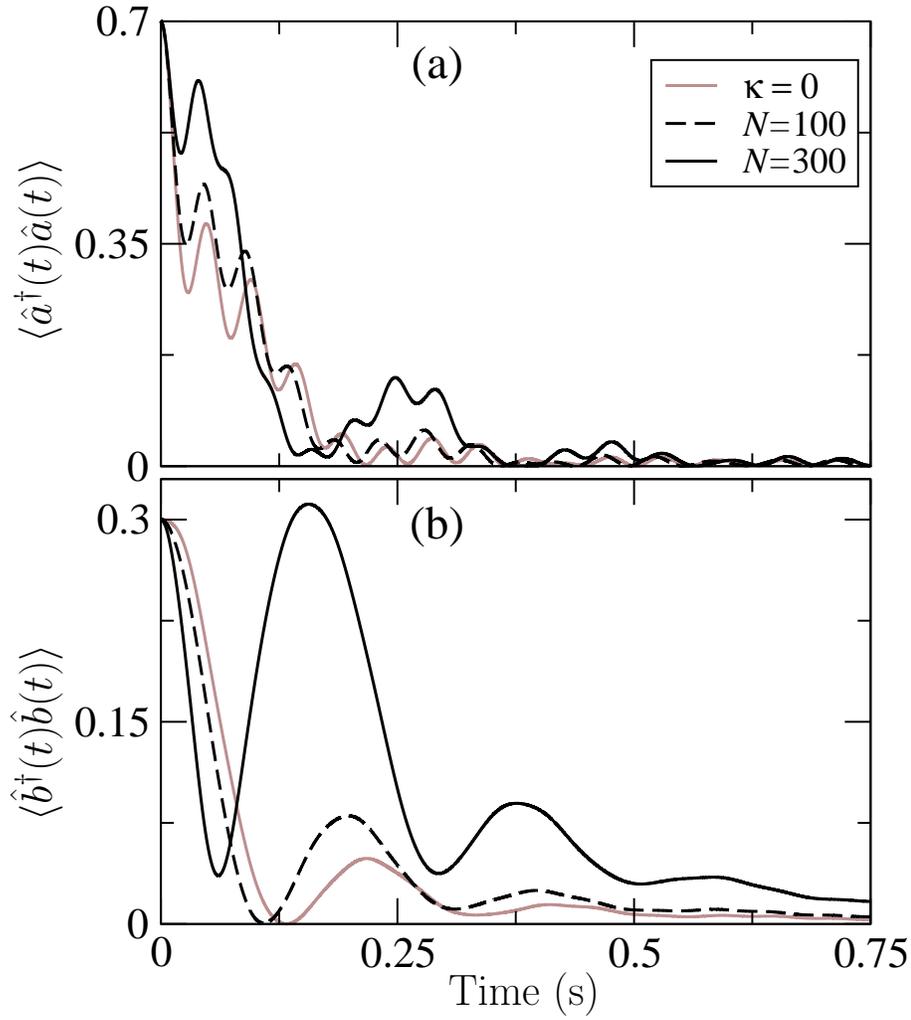}
  \end{center}
\caption{
As in figure \ref{nmarkP1:fig} starting from the self-trapping regime 
for $\Lambda=2\times 10^3~{\rm s}^{-2}$ and $\eta=1.6$. 
} \label{nmarkP2:fig}
\end{figure}

In figures \ref{nmarkP1:fig} and \ref{nmarkP2:fig}, we plot the evolution of 
the occupation probabilities for the 
traps as functions of time. The gray curve refers to an interaction-free 
model whereas the other two curves have been obtained for a weakly interacting 
gas when the system starts from the Josephson and the 
self-trapping regime. 
In any case, when $\xi<1$ we may note only small deviations from the 
interaction-free model (see figure \ref{nmarkP1:fig}). 
The reason is that in the strong-outcoupling regime the evolution of the 
system is mainly governed by the exchange of population between the BECs and
the continuum. Thus, as long as $\xi<1$, the effect of interactions on the 
evolution of the system is negligible, whereas when the system starts from 
the self-trapping regime and $\xi\gtrsim 1$,  we note large deviations 
from the interaction-free model (see black solid line 
in figure \ref{nmarkP2:fig}).  
Moreover, for $N=100$ and the parameters of figure \ref{nmarkP1:fig}, 
$H_{\rm c}(t)<1$ for all times, and thus the system remains in the 
Josephson regime throughout its evolution. 
On the contrary, for $N=200$, as well as for the parameters of figure 
\ref{nmarkP2:fig}, a transition from the self-trapping to the Josephson 
regime occurs rather soon i.e., at $t\approx 9-13~{\rm ms}$.

Our simulations also show a decrease of the steady-state 
population of trap A which is associated with the formation of a bound state 
between the lasing condensate mode and the continuum. 
The destruction of the bound state is due to the repulsive character 
of the interactions and has been predicted earlier by many authors in the 
context of the single-mode model \cite{moypra299,jefpra00}. 
Here, however, we see that the same 
phenomenon also persists in the two-mode model under consideration while, 
as depicted in the inset of figure \ref{nmarkP1:fig}(a), 
the drop of the steady-state population for increasing atom numbers can be 
considered as linear to a good accuracy.

\begin{figure}
  \begin{center}
    \leavevmode
    \epsfxsize12.0cm
    \epsfbox{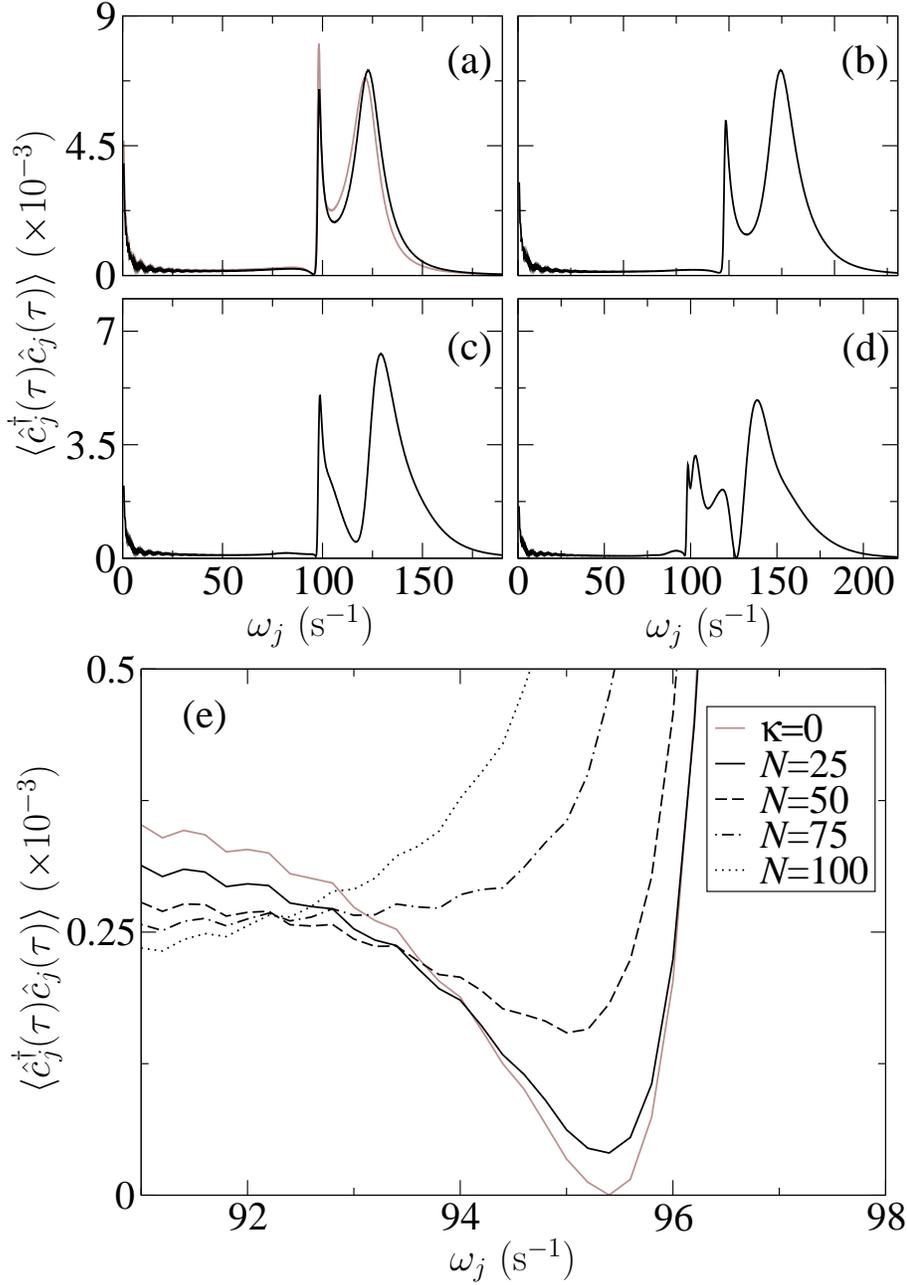}
  \end{center}
\caption{
Effects of interatomic interactions on the distribution of outcoupled atoms 
in the non-Markovian regime. 
(a-d) Atomic distributions at $\tau=10~{\rm s}$ for weakly interacting Bose 
gas with $N=50$ (a),  $N=100$ (b), $N=200$ (c) and $N=300$ (d).
Parameters: $\omega_z=200~{\rm s}^{-1}$, $\lambda=0.4$, 
$\Lambda=2\times 10^3~{\rm s}^{-2}$, $\eta=1.7$. 
(e) A closeup of the atomic distribution around the dip for 
$\Lambda=10^3~{\rm s}^{-2}$. The gray curves correspond to an 
interaction-free Bose gas.
Initial conditions:  
$\tilde{\alpha}(0)=0.7$,  $\tilde{\beta}(0)=0.3$, $\phi(0)=0$. 
Discretization parameters: $M=1500$, $\omega_{\rm up}=300~{\rm s}^{-1}$. 
}  \label{nmarkSpec1:fig}
\end{figure}

The effect of interactions on the distribution of outcoupled atoms 
in the strong-outcoupling regime is summarized in figures \ref{nmarkSpec1:fig}. 
For relatively weak nonlinearities 
(i.e., for $\xi<1$) the distribution of the outcoupled atoms keeps 
all the main features noted in the interaction-free model 
[see figures \ref{nmarkSpec1:fig}(a-b)]. The atomic distributions are mostly affected 
by the presence of interactions when $\xi\gtrsim 1$ 
[see \ref{nmarkSpec1:fig}(c-d)]. 
In this case the chirping effect sets in and the distribution becomes broader. 
Moreover, quantum interference between atoms which have the same frequency but 
were outcoupled at different times may give rise to rich patterns involving 
multiple peaks and dips. Nevertheless, in contrast to the weak-outcoupling regime 
discussed earlier, these peaks (dips) are broader in the present case due to 
the strong outcoupling rates (compare to figure \ref{MarkSpec:fig}). 

Finally, we would like to emphasize once more that the origin of the
dips in the atomic distribution of figure \ref{nmarkSpec1:fig}(c), 
is completely different from the one of the dark spectral line 
predicted in \cite{LazNikLam07}. More precisely, as discussed earlier, 
the multiple dips depicted in figure \ref{nmarkSpec1:fig}(c) stem from the interference between atoms 
which have the same energy but were outcoupled at different times (i.e., they have 
different phases). Interference patterns of this type are not expected to 
depend considerably on the choice of the initial relative phase between the 
two BECs, and thus they are present for any choice of $\phi(0)$.  
On the contrary, the dark line predicted in \cite{LazNikLam07} 
is a signature of perfectly destructive quantum interference between 
different outcoupling channels in the system. This fact was also confirmed 
in section \ref{SNM-phase}, where we showed that the dark line is perfect only 
for $\phi=0,~{\rm and}~ 2\pi$. 
Moreover, as shown in figure  \ref{nmarkSpec1:fig}(e), 
interactions also tend to disturb the interference phenomena between various 
outcoupling channels 
and thus the dark spectral line is not perfect for $\kappa\neq 0$. 
From the mathematical point of view, this is due to the time-dependent 
nonlinearities which render the two condensate modes off-resonant as well as the 
repulsive nature of the interactions. 

\section{Conclusions}
Using a two-mode model, we have studied aspects of the dynamics of a continuous 
atom laser based on the merging of independently formed BECs. 
In particular, we examined effects of interatomic interactions 
and the role of the relative phase between the two BECs on the evolution of the 
trapped populations and the distribution of outcoupled atoms. Our simulations 
were performed in the weak- and the strong-outcoupling regimes.

For the sake of comparison to earlier work \cite{LazNikLam07}, we studied 
the role of the relative phase in the framework of an interaction-free system, 
and showed that it affects considerably the exchange of 
population between the two BECs, as well as the fraction of the 
population which remains trapped in the long-time limit for strong outcoupling 
rates. 
The distribution of outcoupled atoms exhibits always a characteristic doublet, 
with the phase  determining the relative height of the two peaks.  
Moreover, quantum interference between various outcoupling channels in 
the system, may give rise to new peaks or even dark spectral lines in the 
distribution of outcoupled atoms. All of these phenomena, and in particular 
the unconventional form of the spectrum, stem from the presence of the pumping BEC 
in the neighborhood of the lasing condensate. 
More precisely, as long as the outcoupling mechanism is always on during 
the merging process, the transported BEC is also expected to interact  
with the applied electromagnetic radiation before the completion of 
the merging. 
 
In the absence of interactions the energies of the lasing and the 
pumping condensates are constant, whereas in the presence of 
interactions the two condensates are characterized by their chemical 
potentials, which depend on the number of trapped atoms. 
Hence, both of the potentials 
change as atoms are transferred from one condensate mode to the other, 
but mainly 
due to the applied outcoupling mechanism which affects differently 
the population of the two condensates. The two condensate modes 
become thus off-resonant, and this affects mostly the population exchange 
between the two modes. 
Of course, the most noticeable changes occur for nonlinearities whose strength 
is at least comparable to the outcoupling rate and/or the Josephson 
coupling. Moreover, for sufficiently strong interactions 
macroscopic quantum self-trapping may occur, where the 
population transfer from one mode to the other is partially turned off. 
This is a nonlinear effect arising from the interatomic interactions, 
and is preserved in time when the double-well system is isolated. 
In the present case, however, the presence of outcoupling affects the 
trapped populations and the relative phase between the two BECs, 
causing the system to enter gradually the Josephson regime, where it 
remains for the rest of its evolution. 
In the strong-outcoupling regime interatomic interactions seem to prevent 
the formation of a bound state between the lasing mode and the continuum. 
In particular, our simulations show that the steady-state trapped population 
decreases linearly with increasing strength of the nonlinearity. 

Finally, the variation of the chemical potential of the lasing condensate 
with time is also reflected in the spectrum of the atom laser, as 
the outcoupled atoms experience a time-dependent mean-field 
potential, and thus their energy also varies over time (chirping).
This effect causes a significant broadening of the distribution of outcoupled 
atoms, which may also exhibit complicated patterns reflecting the interference 
between atoms which have the same energy but were outcoupled at different 
time instants.
Reducing the effect of the mean-field potential on the outcoupled atoms 
(e.g., by reducing the condensate density or working with interaction-free BEC 
systems in the vicinity of Fesbach resonances) one may in general improve the 
spectrum of the atom laser. It is worth noting, however, 
that an ideal continuous atoms laser working at steady state is not expected to 
suffer by chirping effects as the population of the lasing mode is kept 
constant.

\section*{Acknowledgments}

The work of GMN and PL was supported in part by the EC RTN EMALI 
(contract No. MRTN-CT-2006-035369).

\Figures
\newpage

\end{document}